\documentclass[dvips]{arxstspdf}
\usepackage{dcolumn,stfloats}
\usepackage{graphicx,stfloats}

\volume{25}
\issue{2}
\pubyear{2010}
\firstpage{172}
\lastpage{190}
\doi{10.1214/10-STS327}

\makeatletter
\newcolumntype{d}[1]{D{.}{.}{#1}}
\newcolumntype{a}[1]{D{,}{,\,}{#1}}
\newproclaim{algorithm}{Algorithm}

\newcommand{\cip}{\stackrel{p}{\longrightarrow}}
\newcommand{\cms}{\stackrel{\mathrm{m.s.}}{\longrightarrow}}
\newcommand{\bmH}{\mathbf{H}}
\newcommand{\bmI}{\mathbf{I}}
\newcommand{\bmQ}{\mathbf{Q}}
\newcommand{\bmt}{\mathbf{t}}
\newcommand{\bmu}{\mathbf{u}}
\newcommand{\bmV}{\mathbf{V}}
\newcommand{\bmX}{\mathbf{X}}
\newcommand{\bmY}{\mathbf{Y}}
\newcommand{\bmzero}{\mathbf{0}}
\newcommand{\bmone}{\mathbf{1}}
\newcommand{\bmnu}{\bolds{\nu}}
\newcommand{\bmPsi}{\bolds{\Psi}}
\newcommand{\bmSigma}{\bolds{\Sigma}}
\newcommand{\bmXc}[1]{\mathbf{X}^{(#1)}}
\newcommand{\bmYc}[1]{\mathbf{Y}^{(#1)}}
\newcommand{\ESJD}[1]{S^{2}_{#1}}
\newcommand{\oal}[1]{\overline{\alpha}_{#1}}
\newcommand{\kxd}{k^{(d)}_x}
\newcommand{\kyd}{k^{(d)}_y}
\newcommand{\psibar}{\overline{\psi}}

\renewcommand{\epsilon}{\varepsilon}
\makeatother

\begin{document}
\begin{frontmatter}

\title{The Random Walk Metropolis: Linking Theory and Practice Through a Case Study}
\runtitle{Linking Theory and Practice}

\begin{aug}
\author[a]{\fnms{Chris} \snm{Sherlock}\corref{}\ead[label=e1]{c.sherlock@lancaster.ac.uk}},
\author[b]{\fnms{Paul} \snm{Fearnhead}}
\and
\author[c]{\fnms{Gareth O.} \snm{Roberts}}
\runauthor{C. Sherlock, P. Fearnhead and G. O. Roberts}

\address[a]{Chris Sherlock is Lecturer,
Department of Mathematics and Statistics, Lancaster University,
Lancaster, LA1 4YF, UK \printead{e1}.}
\address[b]{Paul Fearnhead is Professor,
Department of Mathematics and Statistics, Lancaster University,
Lancaster, LA1 4YF, UK.}
\address[c]{Gareth O. Roberts is Professor,
Department of Statistics, University of Warwick, Coventry, CV4 7AL, UK.}

\end{aug}

%
\begin{abstract}
The random walk Metropolis (RWM) is one of the most common
Markov chain Monte Carlo algorithms in practical use today. Its
theoretical properties have been extensively explored for certain
classes of target, and a number of
results with important practical implications have been derived. This
article draws together a selection of new and existing key results and
concepts and describes
their implications. The impact of each new idea on algorithm
efficiency is demonstrated for the practical example of the
Markov modulated Poisson process (MMPP). A reparameterization of
the MMPP which leads to a highly efficient RWM-within-Gibbs
algorithm in certain circumstances is also presented.
\end{abstract}

%
\begin{keyword}
\kwd{Random walk Metropolis}
\kwd{Metropolis--Hastings}
\kwd{MCMC}
\kwd{adaptive MCMC}
\kwd{MMPP}.
\end{keyword}

\end{frontmatter}

\section{Introduction}
\label{sect.introduction}
Markov chain Monte Carlo (MCMC)
algorithms
provide a framework for sampling from
a target random variable with a potentially complicated
probability distribution $\pi(\cdot)$ by generating a Markov chain
$\bmX^{(1)}, \bmX^{(2)}, \ldots$ with stationary distribution
$\pi(\cdot)$.
The single most widely used subclass of MCMC algorithms is based
around the random walk Metropolis (RWM).

Theoretical properties of RWM algorithms for certain special classes
of target have been investigated extensively. Reviews of RWM
theory have, for example, dealt with optimal scaling and posterior shape
(Roberts and Rosenthal, \citeyear{RobertsRosenthal2001}), and
convergence\break (Roberts, \citeyear{Roberts2003}).
This article does not set out to be a comprehensive review of all
theoretical results pertinent to the RWM.
Instead the article reviews and develops specific aspects of
the theory of RWM efficiency in order to tackle an important and
difficult problem: inference for the Markov modulated Poisson process
(MMPP). It
includes sections on RWM within Gibbs, hybrid algorithms, and adaptive
MCMC, as well as optimal scaling, optimal shaping, and convergence.
A strong emphasis is placed on developing an
intuitive understanding of the processes
behind the theoretical results, and then on
using these ideas to improve the implementation.
All of the RWM algorithms described
in this article are tested against datasets arising from
MMPPs. Realized changes in efficiency are
then compared with theoretical predictions.

Observed event times of an MMPP arise from
a Poisson process whose intensity varies with
the state of an unobserved continuous-time Markov chain.
The MMPP has been used to model a wide variety of clustered point
processes, for example, requests for web pages from users of the World
Wide Web
(Scott and Smyth, \citeyear{ScottSmyth2003}),
arrivals of photons from single-molecule fluorescence experiments
(Burzykowski,\break  Szubiakowski and Ryden, \citeyear{Burzykowskietal2003}; Kou, Xie and Liu, \citeyear{Kouetal2005}), and occurrences of a
rare DNA motif along a genome (Fearnhead and Sherlock, \citeyear{FearnheadSherlock2006}).

In common with mixture models
and other hidden Markov models, inference for the MMPP is
greatly complicated by a lack of knowledge of the hidden data. The
likelihood function often possesses many minor modes since the data
might be approximately described by a hidden process with fewer
states. For this same reason the likelihood often does not approach
zero as certain combinations of parameters approach zero and/or
infinity and so
improper priors lead to improper posteriors
(e.g., Sherlock, \citeyear{Sherlock2005}). Further, as with many
hidden data models the
likelihood is invariant under permutation of
the states, and this ``labeling'' problem leads to posteriors with
several equal modes.

This article focuses on generic concepts and techniques
for improving the efficiency of RWM algorithms whatever the
statistical model. The MMPP provides a nontrivial testing ground for
them. All of the RWM algorithms described in this article are tested
against two simulated MMPP datasets with very different
characteristics. This allows us to demonstrate the influence on
performance of
posterior attributes such as shape and orientation near the
mode and lightness or heaviness of tails.

Section \ref{sect.background} introduces RWM algorithms
and then describes theoretical and practical
measures of algorithm efficiency. Next the two main theoretical
approaches to determining efficiency
are described, and the section ends
with a brief overview of the MMPP and a description of the data analyzed
in this article. Section \ref{sect.theory.practice}
introduces a series of concepts which allow potential improvements in
the efficiency of a RWM algorithm. The intuition behind each concept
is described, followed by theoretical justification and then details
of one or more RWM algorithms motivated by the theory. Actual results
are described and compared with theoretical predictions in Section
\ref{sect.results}, and the article is summarized in Section
\ref{sect.discussion}.

\section{Background}
\label{sect.background}
In this section we introduce the background material on which the
remainder of this article draws. We describe the
random walk Metropolis algorithm and
a variation, the random walk Metropolis-within-Gibbs. Both practical
issues and theoretical\break approaches to algorithm efficiency are then
discussed. We conclude with an introduction to the Markov
modulated Poisson process and to the datasets used later in the article.

\subsection{Random Walk Metropolis Algorithms}
\label{sect.genrwm}
The \textit{random walk Metropolis} (RWM) updating\break  scheme
was first applied by Metropolis et al. (\citeyear{Metropolisetal1953}) and proceeds as follows.
Given a current value of the $d$-dimensional Markov chain, $\bmX$,
a new value $\bmX^*$ is obtained by proposing a jump $\bmY^*:=\bmX
^*-\bmX$
from the prespecified Lebesgue density
%
\begin{equation}
\label{eqn.rwm.prop}
\tilde{r} (\mathbf{y}^*;\lambda)
:= \frac{1}{\lambda^d} r\biggl(\frac{\mathbf{y}^*}{\lambda}\biggr) ,
\end{equation}
with $r(\mathbf{y}) = r(-\mathbf{y})$ for all $\mathbf{y}$. Here
$\lambda>0$ governs the
overall size of the proposed jump and (see Section
\ref{sect.optimal.scale}) plays a crucial role in determining the
efficiency of any algorithm.
The proposal is then accepted or rejected according to acceptance probability
%
\begin{equation}
\label{eqn.acc.rate}
\alpha(\mathbf{x},\mathbf{y}^*) = \min\biggl(1, \frac{\pi(\mathbf
{x}+\mathbf{y}^*)}{\pi(\mathbf{x})} \biggr).
\end{equation}
If the proposed value is accepted it becomes the next current value
($\bmX' \leftarrow\bmX+\bmY^*$);
otherwise the current value is left unchanged ($\bmX' \leftarrow\bmX$).

An intuitive interpretation of the above formula is that ``uphill''
proposals (proposals which take the chain closer to a local mode) are
always accepted, whereas ``downhill'' proposals are accepted with
probability exactly equal to the relative ``heights'' of the posterior at
the proposed and current values. It is precisely this rejection of some
``downhill'' proposals which acts to keep the Markov chain in the main
posterior mass most of the time.

More formally, denote by $P(\mathbf{x},\cdot)$ the transition
kernel of the chain, which represents the
combined process of proposal and acceptance/rejection leading from
one element of the chain ($\mathbf{x}$) to the next.
The acceptance probability (\ref{eqn.acc.rate}) is chosen so
that the chain is reversible at equilibrium
with stationary distribution
$\pi(\cdot)$. Reversibility
[that $\pi(\mathbf{x})P(\mathbf{x},\mathbf{x}')=\pi(\mathbf
{x}')P(\mathbf{x}',\mathbf{x})$] is an
important property precisely because it is so easy to construct reversible
chains which have a prespecified stationary distribution. It is also
possible to
prove a slightly stronger central limit theorem for reversible (as
opposed to nonreversible)
geometrically ergodic chains (e.g., Section
\ref{sect.converge}).

We now describe a generalization of the RWM which
acts on a target
whose components have been split into $k$ sub-blocks. In
general we write $\bmX= (\bmX_1, \ldots, \bmX_k
)$, where $\bmX_i$ is the
$i$th sub-block of components of the current element of
the chain.
Starting from value $\bmX$, a single iteration
of this algorithm cycles through all of the sub-blocks updating each in turn.
It will therefore be convenient to define the shorthand
\begin{eqnarray*}
\mathbf{x}^{(B)}_{i} &:=& \mathbf{x}'_1,\ldots,\mathbf
{x}'_{i-1},\mathbf{x}_i,\mathbf{x}_{i+1},\ldots,\mathbf{x}_k,\\
\mathbf{x}^{(B)*}_{i} &:=& \mathbf{x}'_1,\ldots,\mathbf
{x}'_{i-1},\mathbf{x}_i+\mathbf{y}_i^*,\mathbf{x}_{i+1},\ldots
,\mathbf{x}_k ,
\end{eqnarray*}
where $\mathbf{x}'_j$ is the \textit{updated value} of the $j$th sub-block.
For the $i$th sub-block a jump $Y^*_i$ is proposed from symmetric
density $\tilde{r}_i(\mathbf{y};\lambda_i)$ and accepted or rejected
according to acceptance probability
$\pi(\mathbf{x}^{(B)*}_{i} )/\break\pi(\mathbf
{x}^{(B)}_{i} )$.
Since this algorithm is in fact a generalization of both the RWM and
the Gibbs
sampler (for a description of the Gibbs sampler see, e.g.,
 Gamerman and Lopes, \citeyear{GamermanLopes2006}) we follow, for example,
Neal and Roberts (\citeyear{NealRoberts2006}) and call this the
\textit{random walk\break Metropolis-within-Gibbs} or
RWM-within-Gibbs.\break
The most commonly used random
walk\break Metropolis-within-Gibbs algorithm, and also the simplest, is that
employed in this article: here all blocks have dimension 1 so
that each component of the parameter vector is updated in turn.

As mentioned earlier in this section,
the RWM is reversible; but even though each stage of
the RWM-within-Gibbs is reversible, the algorithm as a
whole is not.
Reversible variations
include the \textit{random scan} RWM-within-Gibbs, wherein at each
iteration a
single component is chosen at random and updated conditional on all
the other components.

Convergence of the Markov chain to its stationary distribution can be
guaranteed for all of the above algorithms
under quite general circumstances
(e.g., Gilks, Richardson and Spiegelhalter, \citeyear{GilksRichardsonSpiegelhalter1996}).

\subsection{Algorithm Efficiency}
Consecutive draws of an MCMC Markov chain are correlated and the
sequence of marginal distributions converges to $\pi(\cdot)$.
Two main (and related)
issues arise with regard to the efficiency of MCMC algorithms:
convergence and mixing.

\subsubsection{Convergence}
\label{sect.converge}
In this article we will be concerned with
practical determination of a point at which a chain has
converged. The method we employ is simple heuristic examination of
the trace plots for the different components of the chain. Note that
since the
state space is multidimensional it is not sufficient to simply
examine a single component. Alternative techniques are discussed in
Chapter 7 of the book by Gilks, Richardson and Spiegelhalter (\citeyear{GilksRichardsonSpiegelhalter1996}).

Theoretical criteria for ensuring
convergence (ergodicity) of MCMC Markov chains are examined in
detail in Chapters 3 and 4 of the book by
Gilks, Richardson and Spiegelhalter (\citeyear{GilksRichardsonSpiegelhalter1996}) and references therein,
and will not be discussed
here. We do, however, wish to highlight the
concepts of geometric and polynomial ergodicity.
A Markov chain with transition kernel $P$ is \textit{geometrically ergodic}
with stationary distribution $\pi(\cdot)$ if
%
\begin{equation}
\label{eqn.geom.cvgc.def}
\| P^n(\mathbf{x},\cdot)-\pi(\cdot) \|_1 \le M(\mathbf{x}) r^n
\end{equation}
for some positive $r<1$ and $M(\cdot)\ge0$; if $M(\cdot)$ is bounded
above, then the chain is \textit{uniformly ergodic}.
Here
$\| F(\cdot)-G(\cdot) \|_1$ denotes the total variational distance
between measures $F(\cdot)$ and $G(\cdot)$
(see, e.g., Meyn and Tweedie, \citeyear{MeynTweedie1993}), and $P^n$ is the $n$-step
transition kernel.
Efficiency of a geometrically ergodic algorithm is
measured by the geometric rate of convergence, $r$, which over a large
number of iterations is well approximated by the second largest
eigenvalue of the transition kernel [the largest eigenvalue being $1$,
and corresponding to the stationary distribution
$\pi(\cdot)$]. Geometric ergodicity is usually a purely qualitative
property since in general the constants $M(\mathbf{x})$ and $r$ are
not known.
Crucially for practical MCMC, however, any \textit{geometrically
ergodic reversible Markov chain satisfies a central limit theorem for
all functions with finite second moment with respect to $\pi(\cdot)$}.
Thus there is a $\sigma^2_f < \infty$ such that
%
\begin{equation}
\label{eqn.cent.limit}
n^{1/2} \bigl(\hat{f}_n-\mathbb{E}_{\pi}[f(\bmX)] \bigr)
\Rightarrow
N(0,\sigma_f^2),
\end{equation}
where $\Rightarrow$ denotes convergence in distribution.
The central limit theorem (\ref{eqn.cent.limit}) not only guarantees
convergence of the Monte Carlo estimate (\ref{eqn.mc.average.b})
but also supplies its
standard error, which decreases as $n^{-{1/2}}$.

When the second largest eigenvalue is also $1$, a Markov chain
is termed \textit{polynomially ergodic} if
\[
\| P^n(\mathbf{x},\cdot)-\pi(\cdot) \|_1 \le M(\mathbf{x}) n^{-r}.
\]
Clearly polynomial ergodicity is a weaker condition than geometric
ergodicity. Central limit theorems for polynomially ergodic MCMC are
much more delicate; see the article by Jarner and Roberts (\citeyear{JarnerRoberts2002}) for details.

In
this article a chain is referred to as having ``reached stationarity'' or
``converged'' when
the distribution from which an element is sampled is as close to the
stationary distribution as to make no practical difference to any
Monte Carlo estimates.

An estimate of the expectation of a given function $f(X)$, which is
more accurate than a naive Monte Carlo average over all
the elements of the chain, is
likely to be obtained by discarding the portion of the chain
$\bmX_0,\ldots,\bmX_m$ up until
the point at which it was deemed to have reached stationarity;
iterations
$1,\ldots, m$ are commonly
termed ``burn in.''
Using only the remaining elements
$\bmX_{m+1},\ldots,\bmX_{m+n}$ (with $m+n=N$) our Monte Carlo
estimator becomes
%
\begin{equation}
\hat{f}_n := \frac{1}{n}\sum_{m+1}^{m+n}{f(\bmX_i)}.
\label{eqn.mc.average.b}
\end{equation}
Convergence and burn in are not discussed any further here, and for the
rest of this section
the chain is assumed to have started at stationarity and continued
for $n$ further iterations.

\subsubsection{Practical measures of mixing efficiency}

For a stationary chain,
$\bmX_0$ is sampled from $\pi(\cdot)$, and so for all $k>0$ and
$i\ge0$
\[
\operatorname{Cov}[f(\bmX_k),f(\bmX_{k+i})]=\operatorname
{Cov}[f(\bmX_0),f(\bmX_{i})].
\]
This is the \textit{autocorrelation} at lag $i$. Therefore
at stationarity, from the definition in (\ref{eqn.cent.limit}),
\begin{eqnarray*}
\sigma^2_f &:=& \lim_{n\rightarrow\infty} {n \operatorname{Var}[\hat
{f}_n]}\\ &=&
\operatorname{Var}[f(\bmX_0)] + 2 \sum_{i=1}^\infty{\operatorname
{Cov}[f(\bmX_0),f(\bmX_i)]}
\end{eqnarray*}
provided the sum exists (e.g., Geyer, \citeyear{Geyer1992}).
If elements of the stationary chain were independent, then $\sigma^2_f$
would simply\vspace*{-2pt} be $\operatorname{Var}[f(\bmX_0)]$ and so a measure of
the inefficiency
of the Monte Carlo estimate $\hat{f}_n$ relative to the perfect
i.i.d. sample is
%
\begin{equation} \quad
\frac{\sigma^2_f}{\operatorname{Var}[f(\bmX_0)]}
=
1 + 2 \sum_{i=1}^\infty{\operatorname{Corr}[f(\bmX_0),f(\bmX_i)]}.
\label{eqn.ACT.defn}
\end{equation}
This is the \textit{integrated autocorrelation time} (ACT) and represents
the effective number of dependent samples that is equivalent to a
single independent sample. Alternatively $n^*=n/ACT$ may be regarded as
the effective
equivalent sample size if the elements of the chain had been
independent.

To estimate the ACT in practice one might examine the chain from the
point at which it is deemed to have converged and estimate the lag-$i$
autocorrelation $\operatorname{Corr}[f(\bmX_0),f(\bmX_i)]$ by
%
\begin{equation}
\label{eqn.autocorr.est} \quad
\hat{\gamma}_i =
\frac{1}{n-i}\sum_{j=1}^{n-i}
 \bigl(f(\bmX_j)-\hat{f}_n \bigr) \bigl(f(\bmX_{j+i})-\hat
{f}_n \bigr).
\end{equation}
Naively, substituting these into (\ref{eqn.ACT.defn}) gives an
estimate of the ACT. However, contributions from all terms with very low
theoretical autocorrelation \textit{in a real run}
are effectively random noise, and the sum
of such terms can dominate the deterministic
effect in which we are interested
(e.g., Geyer, \citeyear{Geyer1992}).
For this article we employ the simple solution suggested by
Carlin and Louis (\citeyear{CarlinLouis2009}):
the sum (\ref{eqn.ACT.defn}) is truncated from the first lag, $l$,
for which the estimated autocorrelation
drops below $0.05$. This gives the (slightly biased) estimator
%
\begin{equation}
\operatorname{ACT}_{\mathrm{est}} :=
1 + 2 \sum_{i=1}^{l-1}{\hat{\gamma}_i}.
\label{eqn.ACT.est.defn}
\end{equation}
Given the potential for relatively large variance in estimates of
integrated ACT howsoever they might be obtained
(e.g., Sokal, \citeyear{Sokal1996}), this simple estimator
should be adequate for comparing the relative efficiencies of
the different algorithms in this article.
Geyer (\citeyear{Geyer1992}) provided a number of more complex window estimators
and provided references for regularity conditions under
which they are consistent.

A given run will have a different ACT associated with each
parameter. An alternative efficiency measure, which is aggregated over
all parameters, is provided by the Mean Squared Euclidean Jump Distance (MSEJD)
\[
S^2_{\mathrm{Euc}} := \frac{1}{n-1}\sum_{i=1}^{n-1}{\bigl\| \mathbf
{x}^{(i+1)}-\mathbf{x}^{(i)} \bigr\|_2^2}.
\]
The expectation of this quantity at stationarity is referred to as the
Expected Squared Euclidean Jump Distance (ESEJD).
Consider a single component of the target with variance
$\sigma_i^2 := \operatorname{Var}[X_i] =
\operatorname{Var}[X_i']$, and note that $\mathbb{E}[X_i'-X_i]=0$, so
\begin{eqnarray*}
\mathbb{E}[(X_i'-X_i)^2] &=& \operatorname{Var}[X_i'-X_i]\\ &=&
2 \sigma_i^2 (1-\operatorname{Corr}[X_i,X_i'] ).
\end{eqnarray*}
Thus when the chain is stationary and the posterior variance is finite,
maximizing the ESEJD is equivalent to
minimizing a weighted sum of the lag-1 autocorrelations.

If the target has finite second moments and is roughly elliptical in
shape with (known) covariance matrix $\bmSigma$, then an alternative
measure of efficiency is the Mean Squared Jump Distance (MSJD)
\[
S^2_d := \frac{1}{n-1}\sum_{i=1}^{n-1}
 \bigl(\mathbf{x}^{(i+1)}-\mathbf{x}^{(i)} \bigr)^t\bmSigma
^{-1} \bigl(\mathbf{x}^{(i+1)}-\mathbf{x}^{(i)} \bigr),
\]
which is proportional to the unweighted sum of the \mbox{lag-1}
autocorrelations over the principal components of the ellipse.
The theoretical expectation of the MSJD at stationarity is known as
the expected\break squared jump distance (ESJD).

\begin{figure*}

\includegraphics{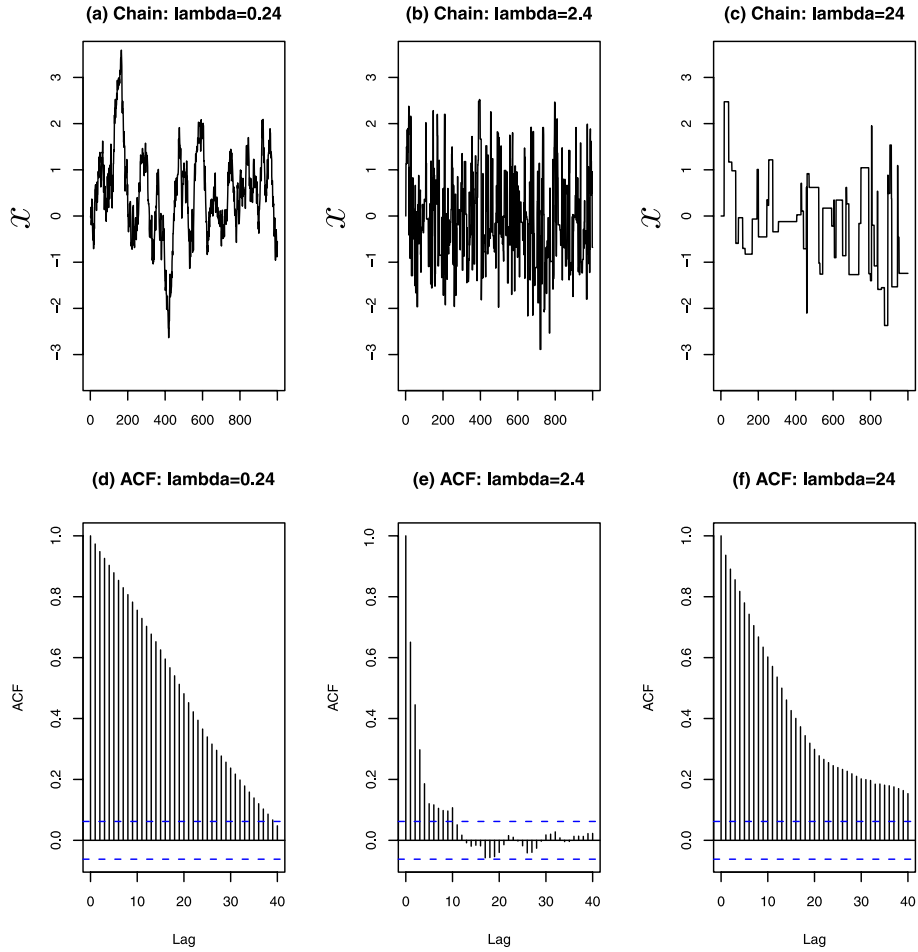}

\caption{Traceplots [(\textup{a}), (\textup{b}), and (\textup{c})] and corresponding
autocorrelation plots
[(\textup{d}), (\textup{e}), and (\textup{f})],
for exploration of a standard Gaussian initialized from
$x=0$ and using the random
walk Metropolis algorithm with Gaussian proposal for 1000 iterations.
Proposal scale
parameters for the three scenarios are, respectively, (\textup{a}) and (\textup{d})
$0.24$, (\textup{b}) and (\textup{e}) $2.4$,
and (\textup{c}) and (\textup{f}) $24$.}
\label{fig.simple.traceplots}
\end{figure*}

Figure \ref{fig.simple.traceplots} shows traceplots for three
different\break Markov chains. Estimates of the autocorrelation from lag-0
to lag-40 for each Markov chain appear alongside the corresponding
traceplot. The simple window estimator for
integrated ACT provides estimates of,
respectively, $39.7$, $5.5$, and $35.3$. The MSEJDs are,
respectively, $0.027$, $0.349$, and $0.063$, and are equal to the
MSJDs since the stationary distribution has a variance of $1$.\looseness=-1

\subsubsection{Assessing accuracy}
\label{sect.accuracy}
An MCMC algorithm\break might efficiently explore an unimportant part
of the parameter space and never find the main posterior mass.
ACT's will be low, therefore, but the resulting
posterior estimate will be wildly inaccurate. In most practical
examples it is not possible to determine the accuracy of the posterior
estimate, though consistency between several independent runs or
between different portions of the same run can be tested.

For the
purposes of this article it was important to have a relatively accurate
estimate of the posterior, not determined by a RWM
algorithm. Fearnhead and Sherlock (\citeyear{FearnheadSherlock2006}) detailed a Gibbs sampler for
the MMPP; this Gibbs sampler
was run for 100,000 iterations on each of the datasets
analyzed in this article. A ``burn in'' of 1000 iterations was allowed
for, and a posterior estimate from the last 99,000
iterations was used as a reference for comparison with
posterior estimates from RWM runs of 10,000 iterations (after burn in).

\subsubsection{Theoretical approaches for algorithm\break efficiency}
\label{sect.scaling.theory}
To date, theoretical results on the efficiency of RWM algorithms have
been obtained\break through two very different approaches. We
wish to quote, explain, and apply theory from both and so we give
a heuristic description of each and define associated
notation. Both approaches link some measure of efficiency to the
expected acceptance rate---the expected proportion of proposals
accepted at stationarity.

The first approach was pioneered by
Roberts, Gelman and Gilks (\citeyear{RobertsGelmanGilks1997}) for targets with independent
identically distributed components and then generalized by
Roberts and Rosenthal (\citeyear{RobertsRosenthal2001}) to targets of the form
\[
\pi(\mathbf{x}) = \prod_1^d{C_i f(C_i x_i)}.
\]
The inverse scale parameters, $C_i$, are assumed to be
drawn from some distribution with a given (finite)
mean and variance.
A single component of the $d$-dimensional chain
(without loss of generality the first)
is then examined; at iteration $i$ of the algorithm it is denoted
$X_{1,i}^{(d)}$.
A scaleless, speeded up,
continuous-time process which mimics the first component of the chain
is defined as
\[
W_{t}^{(d)} := C_1 X^{(d)}_{1,[td]},
\]
where $[u]$ denotes the nearest integer less than or equal to $u$.
Finally, proposed jumps are assumed to be Gaussian
\[
\bmY^{(d)} \sim N (\bmzero,\lambda_d^2\bmI).
\]
Subject to conditions on the first two deriviatives of
$f(\cdot)$, Roberts and Rosenthal (\citeyear{RobertsRosenthal2001}) showed that if $\mathbb
{E}[C_i]=1$ and
$\mathbb{E}[C_i^2]=b$, and provided $\lambda_d = \mu/d^{1/2}$ for some
fixed $\mu$
(the scale parameter but ``rescaled'' according to dimension),
then as $d\rightarrow\infty$, $W_t^{(d)}$ approaches a Langevin
diffusion process with speed
%
\begin{eqnarray}
\label{eqn.diff.speed.acc}
h(\mu) = \frac{C_1^2 \mu^2}{b} \oal{d}\nonumber
\\[-8pt]
\\[-8pt]
\eqntext{\mbox{where }
\oal{d} := 2 \Phi\biggl(-\dfrac{1}{2}\mu J^{1/2} \biggr).}
\end{eqnarray}
Here $\Phi(x)$ is the cumulative distribution function of a standard
Gaussian, $J := \mathbb{E}[((\log f)' )^2]$ is a measure of the
roughness of the target, and $\oal{d}$
corresponds to the acceptance rate.

B{\'e}dard (\citeyear{BedardA2007}) proved a similar result for a triangular
sequence of inverse scale parameters $c_{i,d}$, which are assumed to
be known. A
necessary and sufficient condition
equivalent to (\ref{eqn.suff.cond}) below is attached to this result.
In effect this requires the scale over which the smallest component
varies to be ``not too much smaller'' than the scales of the other components.

The second technique (e.g., Sherlock and Roberts, \citeyear{SherlockRoberts2009}) uses
expected squared jump distance (ESJD) as a measure of efficiency.
Exact analytical forms for ESJD (denoted $\ESJD{d}$)
and expected acceptance rate are
derived for any unimodal elliptically symmetric target and any
proposal density. Many standard sequences of
$d$-dimensional targets ($d=1,2,\ldots$), such as the Gaussian, satisfy
the condition that as $d
\rightarrow\infty$ the probability mass becomes concentrated in a
spherical shell which itself becomes infinitesimally thin relative to its
radius. Thus the random walk on a rescaling of the target is, in the limit,
effectively confined to the surface of this shell.
Sherlock and Roberts (\citeyear{SherlockRoberts2009}) considered
a sequence of targets which satisfies such a ``shell'' condition, and
a sequence of proposals which satisfies a slightly
stronger condition. Specifically it is required that there exist
sequences of positive real numbers, $\{\kxd\}$ and $\{\kyd\}$, such that
\[
\frac{\| \bmXc{d} \|}{\kxd} \cip1
 \quad \mbox{and} \quad
\frac{\| \bmYc{d} \|}{\lambda_d \kyd} \cms1.
\]
For such combinations of target and proposal,
as $d \rightarrow\infty$
%
\begin{eqnarray}\label{eqn.sphlimitsa.cip.ell}
\frac{d}{{\kxd}^2}\ESJD{d}(\mu) \rightarrow
\mu^2 \oal{d}\nonumber
\\[-8pt]
\\[-8pt]
\eqntext{\mbox{with }
\oal{d}(\mu) :=
2 \Phi\biggl(-\dfrac{1}{2}\mu\biggr).}
\end{eqnarray}
Here $\oal{d}$ is the limiting expected acceptance rate, and
$\mu:= d^{1/2} \lambda_d \kyd/\kxd$.
For target and proposal
distributions with independent components, such as are used in the
diffusion results, $\kxd=\kyd=d^{1/2}$, and hence (consistently) $\mu
= d^{1/2}
\lambda_d$.

It is also required that the elliptical target not be too eccentric.
Specifically,
for a sequence of target densities
$\pi_d(\mathbf{x}):=f_d (\sum_{i=1}^dc_{i,d}^2x_i^2 )$
(for some
appropriate sequence of functions $\{f_d\}$)
%
\begin{equation}
\label{eqn.suff.cond}
\frac{\max_{i} c_{i,d}^2}{\sum_{i=1}^d{c_{i,d}^2}}
\rightarrow0  \quad \mbox{as } d\rightarrow\infty.
\end{equation}

Theoretical results from the two techniques are remarkably similar and
as will be seen, lead to identical strategies for optimizing algorithm
efficiency. It is worth noting, however, that results from
the first approach apply
only to targets with independent components and results from the second
only to
targets which are unimodal and elliptically symmetric. That they lead
to identical strategies indicates a certain potential robustness of these
strategies to the form of the target. This potential, as we shall see,
is borne
out in practice.

\subsection{The Markov Modulated Poisson Process}
Let $X_t$ be a continuous-time Markov chain on discrete state space
$\{1,\ldots,d\}$ and let $\bolds{\psi}:=[\psi_1,\ldots
,\psi_d]$ be a
$d$-dimensional vector of
(nonnegative) intensities. The linked but stochastically independent
Poisson process $Y_t$ whose intensity
is $\psi_{X_t}$ is a Markov modulated Poisson process---it is a
Poisson process whose intensity is modulated by a continuous-time
Markov chain.

The idea is best illustrated through two examples, which also serve to
introduce the notation and datasets
that will be used throughout this article.
Consider a two-dimensional Markov chain $X_t$ with generator $\bmQ$
with $q_{12}=q_{21}=1$.

\begin{figure}

\includegraphics{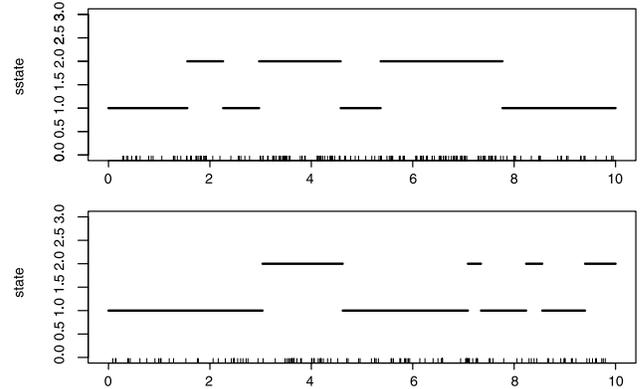}

\caption{Two 2-state
continuous-time Markov chains simulated for 10 seconds from generator
$\bmQ$ with $q_{12}=q_{21}=1$; the rug plots show events from
an MMPP simulated from these chains, with intensity vectors
$\bolds{\psi}=[10,30]$ (upper graph) and $\bolds{
\psi}=[10,17]$ (lower graph).}
\label{fig.mmpp.example}
\end{figure}

Figure \ref{fig.mmpp.example} shows realizations from two such
chains over a period of $10$ seconds. Now consider a Poisson process
$Y_t$ which
has intensity $10$ when $X_t$ is in state~$1$ and intensity $30$ when
$X_t$ is in state $2$. This is an MMPP with event intensity vector
$\bolds{\psi}= [10,30 ]$.
A realization (obtained via the realization of $X_t$)
is shown as a rug plot underneath the chain in the upper graph. The
lower graph shows a realization from an MMPP with event intensities
$[10,17]$.

It can be shown (e.g., Fearnhead and Sherlock, \citeyear{FearnheadSherlock2006}) that the
likelihood for data from an MMPP which starts from a
distribution $\bmnu$ over its states is
%
\begin{eqnarray}
&&L( \bmQ, \bmPsi, \bmt)\nonumber
\\[-8pt]
\\[-8pt] && \quad =
\bmnu' e^{(\bmQ- \bmPsi)t_1} \bmPsi\cdots e^{(\bmQ-
\bmPsi)t_n}\bmPsi
e^{(\bmQ- \bmPsi)t_{n+1}}\bmone.\nonumber
\label{eqn.etlikelihood}
\end{eqnarray}
Here $\bmPsi:=\operatorname{diag}(\bolds{\psi})$,
$\bmone$ is a vector of $1$'s, $n$ is the number of observed events,
$t_1$ is the time from the start of the observation window until the first
event, $t_{n+1}$ is the time from the last event until the end of the
observation window, and $t_{i} (2\le i\le n)$
is the time between the $(i-1)$th and $i$th events. In the absence
of further information, the initial
distribution $\bmnu$ is often taken to be the stationary distribution of
the underlying Markov chain.

The likelihood of an MMPP is invariant to a relabeling of the
states. Hence if the prior is similarly invariant, then so too is the
posterior: if the posterior for a two-dimensional MMPP has a mode at
$(\psi_1,\psi_2,q_{12},q_{21})$, then it has an identical mode
at $(\psi_2,\psi_1,q_{21},q_{12})$. In this article
our overriding interest is in the efficiency of the MCMC algorithms
rather than the exact meaning of the parameters and so we choose the
simplest solution to this identifiability problem: the state
with the lower Poisson intensity $\psi$ is always referred to as
state~1.

\subsubsection{MMPP data in this article}
\label{sect.mmpp.data}
\begin{figure}

\includegraphics{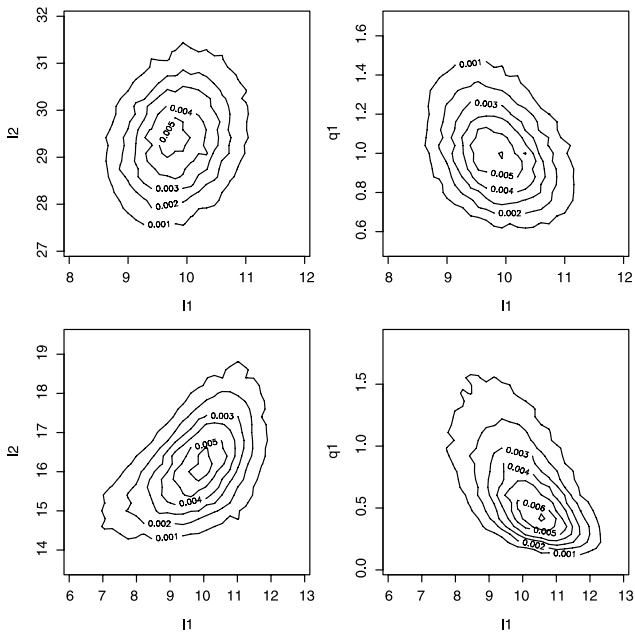}

\caption{Estimated marginal posteriors for $\psi_1$ and
$\psi_2$ and for $\psi_1$ and $q_{12}$ from long runs of the
Gibbs sampler for datasets \textup{D}1 (top) and \textup{D}2 (bottom).}
\label{fig.posterior}
\end{figure}

The two datasets of event times used in this article arose from two
independent MMPP's
simulated over an observation window of 100 seconds.
Both underlying Markov chains have $q_{12}=q_{21}=1$; dataset D1 has
event intensity vector $\bolds{\psi}=[10,30]$ whereas
dataset D2 has
$\bolds{\psi}=[10,17]$, so that the overall intensity of
events in D2 is
lower than in D1. As mentioned in Section \ref{sect.accuracy},
a posterior sample from a long run of the Gibbs
sampler of Fearnhead and Sherlock (\citeyear{FearnheadSherlock2006}) was used to approximate
the true posterior. Figure \ref{fig.posterior} shows estimates of the
marginal
posterior distribution for $(\psi_1,\psi_2)$ and for $(\psi
_1,q_{12})$ for D1 (top) and D2 (bottom).

Because the difference in intensity between
the states is so much larger in D1 than in D2 it is easier with
D1 than D2 to
distinguish the state of the underlying Markov chain, and thus the
values of the Markov and Poisson parameters. Further, in the limit of
the underlying chain being known precisely, for example as
$\psi_2\rightarrow\infty$ with $\psi_1$ finite,
and provided the priors are independent,
the posteriors for the Poisson intensity
parameters $\psi_1$ and $\psi_2$ are completely independent of
each other and of the Markov parameters $q_{12}$ and $q_{21}$.
Dependence between the
Markov parameters is also small, being $O(1/T)$
(e.g., Fearnhead and Sherlock, \citeyear{FearnheadSherlock2006}).

In Section \ref{sect.results},
differences between D1 and D2 will be related directly to
observed differences in efficiency of the various RWM algorithms
between the two datasets.

\section{Implementations of the RWM: Theory and Practice}
\label{sect.theory.practice}
This section describes several theoretical
results for the RWM or for MCMC in general.
Intuitive explanation of the principle behind each result is
emphasized and the manner in which it informs the RWM implementation
is made clear. Each algorithm was run three times on each of
the two datasets.

\subsection{Optimal Scaling of the RWM}
\label{sect.optimal.scale}
\textit{Intuition:} Consider the behavior of the RWM as a function of
the overall scale parameter of the proposed jump, $\lambda$,
in (\ref{eqn.rwm.prop}).
If most proposed jumps are small
compared with some measure of the scale of variability of the
target distribution,
then, although these jumps will often be accepted, the chain will move
slowly and exploration of the target distribution will be relatively
inefficient. If the jumps proposed are relatively large compared
with the target distribution's scale, then many will not be
accepted, the chain will rarely move, and will again explore the
target distribution inefficiently. This suggests that given a
particular target and form for the jump proposal distribution,
there may exist a finite scale parameter for the proposal with which the
algorithm will explore the target as efficiently as possible. These
ideas are clearly demonstrated in Figure
\ref{fig.simple.traceplots} which
shows traceplots for a one-dimensional Gaussian target
explored using a Gaussian proposal with scale parameter an order of
magnitude smaller (a) and larger (c) than is optimal, and (b) with
a close to optimal scale parameter.

\textit{Theory:} Equation (\ref{eqn.diff.speed.acc}) gives algorithm
efficiency for a target with independent and identical (up to a
scaling) components as a function
of the ``rescaled'' scale parameter $\mu= d^{1/2}\lambda_d$ of a
Gaussian proposal. Equation~(\ref{eqn.sphlimitsa.cip.ell}) gives algorithm efficiency for a unimodal
elliptically symmetric target explored by a spherically symmetric
proposal with $\mu= d^{1/2}\lambda_d \kyd/\break\kxd$. Efficiencies are
therefore optimal at $\mu\approx
2.38/\break J^{1/2}$ and $\mu\approx2.38$, respectively.
These correspond to actual scale parameters of respectively
\[
\lambda_d = \frac{2.38}{J^{1/2}d^{1/2}}
 \quad \mbox{and} \quad
\lambda_d = \frac{2.38 \kxd}{d^{1/2}\kyd}.
\]
The equivalence between these two expressions for Gaussian data explored
with a Gaussian target is clear from Section
\ref{sect.scaling.theory}. However,
the equations offer little direct help in choosing a scale parameter
for a target which is neither elliptical nor possesses components which are
i.i.d. up to a scale parameter. Substitution of
each expression into the corresponding acceptance rate equation, however,
leads to the same
optimal acceptance rate, $\hat{\alpha} \approx0.234$. This justifies
the relatively well-known adage that \textit{for random walk
algorithms with a
large number of parameters, the scale parameter of the proposal should
be chosen so that the acceptance rate is approximately $0.234$}.
On a graph of asymptotic efficiency against acceptance rate
(e.g., Roberts and Rosenthal, \citeyear{RobertsRosenthal2001}), the
curvature near the mode is slight, especially to its right,
so that an acceptance rate of
anywhere between $0.2$ and $0.3$ should lead to an algorithm of
close to optimal efficiency.

In practice updates are performed on a finite number of parameters;
for example, a two-dimensional MMPP has four parameters
($\psi_1,\psi_2,q_{12},q_{21}$). A block update involves all
of these, while each update of a
simple Metropolis-within-Gibbs step involves just one parameter. In
finite dimensions the optimal acceptance rate can in fact take any
value between $0$ and $1$. Sherlock and Roberts (\citeyear{SherlockRoberts2009}) provided
analytical formulas for calculating the ESJD
and the expected acceptance rate for any proposal and any
elliptically symmetric unimodal target. In one dimension, for example, the
optimal acceptance rate for a Gaussian target explored by a Gaussian
proposal is $0.44$,
while the optimum for a Laplace
target ($\pi(x) \propto e^{-\vert x \vert}$) explored with a Laplace
proposal is exactly $\hat{\alpha}=1/3$.
Sherlock (\citeyear{Sherlock2006}) considered
several simple examples of spherically symmetric
proposal and target across a range of dimensions
and found that in all cases curvature at the optimal acceptance rate
is small, so that a range of acceptance rates is nearly optimal.
Further, the optimal acceptance rate is itself between $0.2$ and
$0.3$ for $d \ge6$ in all the cases considered.

Sherlock and Roberts (\citeyear{SherlockRoberts2009}) also weakened the ``shell'' condition of
Section \ref{sect.scaling.theory} and considered sequences of
spherically symmetric targets for which the (rescaled)
radius
converges to some random variable $R$ rather than a point mass at
$1$.
It is shown that,
provided the sequence of proposals still satisfies the shell condition,
the limiting optimal acceptance rate is strictly less than $0.234$.
Acceptance rate tuning should
thus be seen as only a guide, though a guide which has been found to be
robust in practice.

\begin{algorithm}[(Blk)]\label{algo1}
The first algorithm (Blk) used to explore datasets D1 and D2
is a four-dimen\-sional block updating RWM with proposal
$\bmY\sim N(0,\break \lambda^2\bmI)$ and
$\lambda$ tuned so that the acceptance rate is approximately $0.3$.
\end{algorithm}

\subsection{Optimal Scaling of the RWM-Within-Gibbs}
\textit{Intuition:} Consider first a target either spherically
symmetric, or with i.i.d. components, and let the overall
scale of variability of the target be $\eta$. For full block proposals
the optimal scale parameter should be $O (\eta/d^{1/2} )$
so that the square of the magnitude of
the total proposal is $O(\eta^2)$.
If a Metropolis-within-Gibbs update is to be used with $k$ sub-blocks and
$d_*=d/k$ of the components
updated at each stage,
then the optimal scale parameter should be larger,
$O (\eta/d_*^{1/2} )$.
However, only one of the $k$ stages of the RWM-within-Gibbs
algorithm updates any given component whereas with $k$ repeats of a
block RWM that component is updated $k$ times. Considering the squared
jump distances it is easy to see that, given the additivity of
squared jump distances, the larger size of the
RWM-within-Gibbs updates is exactly canceled by their lower frequency,
and so (in the limit) there is no difference in efficiency when
compared with a block update. The same intuition applies when
comparing a random scan Metropolis-within-Gibbs scheme with a single
block update.

Now consider a target for which different components vary on different
scales. If sub-blocks are chosen so as to group together components with
similar scales, then a Metropolis-within-Gibbs scheme can apply
suitable scale paramaters to each block whereas a single block update
must choose one scale parameter that is adequate for all
components. In this scenario,
Metropolis-within-Gibbs updates should therefore be more
efficient.

\textit{Theory:} Neal and Roberts (\citeyear{NealRoberts2006}) considered a random scan
RWM-within-Gibbs algorithm
on a target distribution with i.i.d. components and using i.i.d.
Gaussian proposals all having the same
scale parameter $\lambda_d = \mu/d^{1/2}$. At each iteration a
fraction, $\gamma_d$, of the $d$ components are chosen uniformly at
random and
updated as a block.
It is shown [again subject to
differentiability conditions on $f(\cdot)$] that the process
$W_t^{(d)}:=X^{(d)}_{1,[td]}$ approaches a Langevin diffusion with
speed
\[
h_\gamma(\mu) = 2\gamma\mu^2\Phi\bigl(-\tfrac{1}{2}\mu(\gamma
J)^{1/2} \bigr),
\]
where $\gamma:=\lim_{d\rightarrow\infty}\gamma_d$.
The optimal scaling is therefore larger than for a standard block
update (by a factor of $\gamma^{-1/2}$) but the optimal speed and the
optimal acceptance rate (0.234) are identical to those found by
Roberts, Gelman and Gilks (\citeyear{RobertsGelmanGilks1997}).

Sherlock (\citeyear{Sherlock2006}) considered sequential Metropolis-within-Gibbs
updates on a unimodal elliptically
symmetric target, using spherical proposal distributions but
allowing \textit{different} scale parameters for the proposals in each
sub-block.
The $k$ sub-blocks are assumed to correspond to disjoint subsets of the
principal axes of the ellipse and updates for each are assumed to be
optimally tuned. Efficiency is considered in terms of ESEJD and is
again found to be optimal (as $d \rightarrow\infty$)
when the acceptance rate for each sub-block is $0.234$. For
equal sized sub-blocks, the relative efficiency of the
Metropolis-within-Gibbs\break
scheme compared to $k$ optimally scaled single block updates is shown
to be\looseness=1
%
\begin{equation}
\label{eqn.MwG.efficiency}
r=
\frac{ ({1}/{k})\sum{\overline{c^2}_i}}
{ ( ({1}/{k})\sum {1}/{\overline{c^2}_i} )^{-1}} ,
\end{equation}
where $\overline{c^2}_i$ is the mean of the squares of the
inverse scale parameters
for the $i$th block. Since $r$ is the ratio of an arithmetic mean
to a harmonic mean, it is greater than or equal to 1 and thus
the Metropolis-within-Gibbs step is always at
least as efficient as the block Metropolis. However, the more similar the
blocks, the less the potential gain in efficiency.

In practice, parameter blocks do not generally correspond to disjoint
subsets of the principal axes of the posterior or, in terms of single
parameter updates, the parameters are not generally
orthogonal. Equation (\ref{eqn.MwG.efficiency}) therefore corresponds
to a limiting
maximum efficiency gain,
obtainable only when the parameter sub-blocks are orthogonal.

\begin{algorithm}[(MwG)]\label{algo2}
Our second algorithm (MwG) is a sequential Metropolis-within-Gibbs
algorithm with proposed jumps $Y_i \sim N(0,\lambda_i^2)$. Each
scale parameter is tuned separately to give an acceptance rate of
between $0.4$ and $0.45$ (approximately the optimum for a
one-dimensional Gaussian
target and proposal).
\end{algorithm}

\subsection{Tailoring the Shape of a Block Proposal}
\textit{Intuition:}
First consider a two-dimensional target with roughly
elliptical contours and with the scale of variation along one of the
principal axes much larger than the scale of variation along the
other (e.g., the two right-hand panels of Figure \ref{fig.posterior}).
The size of updates from a proposal of the type used in
Algorithm~\ref{algo1} is constrained by the smaller of the two
scales of variation. Thus, even when Algorithm~\ref{algo1} is optimally tuned,
the efficiency of exploration along the larger axis depends on the
ratio of the two scales and so can be arbitrarily low in targets where
this ratio is large.
Now consider a general target with roughly elliptical contours and
covariance matrix $\bmSigma$. It seems intuitively sensible
that a ``tailored'' block
proposal distribution with the same shape and orientation as
the target will tend to produce larger jumps along the target's major
axes and smaller jumps along its minor axes and should therefore
allow for more
efficient exploration of the target.

\textit{Theory:}
Sherlock (\citeyear{Sherlock2006}) considered exploration of a unimodal elliptically
symmetric target with either a spherically symmetric proposal or a
tailored elliptically symmetric proposal in the limit as $d
\rightarrow\infty$. Subject to condition
(\ref{eqn.suff.cond}) (and a ``shell''-like condition similar to that
mentioned in Section \ref{sect.scaling.theory}),
it is shown that with each proposal shape
it is in fact possible to achieve the same
optimal expected squared jump distance. However,
if a spherically
symmetric proposal is used on an elliptical target,
some components are explored better
than others and in some sense
the overall efficiency is reduced. This becomes clear
on considering the ratio, $r$, of the expected
squared Euclidean jump distance for an optimal spherically symmetric
proposal to that of an optimal tailored
proposal. Sherlock (\citeyear{Sherlock2006}) showed that for a sequence of targets,
where the target with dimension $d$
has elliptical axes with inverse scale parameters
$c_{d,1},\ldots,c_{d,d}$, the limiting ratio is
\[
r =
\frac{
\lim_{d\rightarrow
\infty} ({({1}/{d})\sum_{i=1}^d{c_{d,i}^{-2}}} )^{-1}}
{\lim_{d\rightarrow\infty}{ ({1}/{d})\sum_{i=1}^d{c_{d,i}^{2}}}}.
\]
The numerator is the limiting harmonic mean of the squared
inverse scale parameters,
which is less than or equal to their arithmetic mean (the denominator), with
equality if and only if (for a given $d$)
all the $c_{d,i}$ are equal.
Roberts and Rosenthal (\citeyear{RobertsRosenthal2001}) examined similar relative efficiencies
but for targets and proposals with independent components with inverse
scale parameters $C$ sampled from some distribution. In this
case the derived measure of relative efficiency
is the relative speeds of the diffusion limits for the
first component of the target
\[
r^* = \frac{\mathbb{E}[C]^2}{\mathbb{E}[C^2]}.
\]
This is again less than or equal to 1, with equality when
all the scale parameters are equal.
Hence efficiency is indeed directly related to the relative
compatibility between target and proposal shapes.

Furthermore, B{\'e}dard (\citeyear{Bedard2008}) showed that if a proposal has
i.i.d. components yet the target (assumed to have independent components)
is wildly asymmetric, as measured by
(\ref{eqn.suff.cond}), then the limiting optimal acceptance rate can be
anywhere between $0$ and $1$. However, even at this optimum, some
components will be explored infinitely more slowly than others.

In practice the shape $\bmSigma$ of the posterior is not known and
must be
estimated, for example by numerically finding the posterior mode and
the Hessian matrix $\bmH$ at the mode, and setting $\bmSigma=
\bmH^{-1}$. We employ a simple alternative which uses an earlier MCMC
run.

\begin{algorithm}[(BlkShp)]\label{algo3} Our third algorithm first uses an
optimally scaled block RWM algorithm (Algorithm~\ref{algo1}), which is
run for long enough to obtain a ``reasonable'' estimate of the
covariance
from the posterior sample. A fresh run is then started and tuned to
give an acceptance rate of about $0.3$ but using proposals
\[
\bmY\sim N(\bmzero, \lambda^2 \hat{\bmSigma}).
\]
For each dataset,
so that our implementation would reflect likely statistical practice,
each of the three replicates of this algorithm estimated the $\bmSigma
$ matrix
from iterations 1000--2000 of the corresponding replicate of Algorithm~\ref{algo1}
 (i.e., using 1000 iterations after ``burn in''). In all, therefore,
six different variance matrices were used.
\end{algorithm}

\subsection{Improving Tail Exploration}
\label{sect.tail.explore}
\textit{Intuition:} A posterior with relatively heavy
polynomial tails such as the one-dimensional Cauchy
distribution has considerable mass some distance
from the origin. Proposal scalings which efficiently explore the
body of the posterior are thus too small to explore much of the tail
mass in
a ``reasonable'' number of iterations. Further, polynomial tails
become flatter with distance from the origin so that for unit vector
$\bmu$,
$\pi(\mathbf{x}+\lambda\bmu)/\pi(\mathbf{x}) \rightarrow1$ as
$\| \mathbf{x} \|_2\rightarrow
\infty$. Hence the acceptance rate for a random walk algorithm
approaches $1$ in the tails, whatever the direction of the proposed jump.
The algorithm therefore loses almost all sense of
the direction to the posterior mass.

\textit{Theory:} Roberts (\citeyear{Roberts2003}) brought together literature
relating the tails of the $d$-dimensional posterior and proposal
to the ergodicity of
the Markov chain and hence its convergence properties. Three important
cases are noted:
\begin{enumerate}[(3)]
\item[(1)]
If $\exists s>0$ such that $\pi(\mathbf{x}) \propto e^{-s\| \mathbf
{x} \|_2}$, at least outside some compact
set, then the random walk algorithm is geometrically ergodic.
\item[(2)]
If $\exists r>0$ such that the tails of the proposal are bounded by
some multiple of
$\| x \|_2^{-(r+d)}$ and if $\pi(\mathbf{x}) \propto\| \mathbf {x}
\|_2^{-(r+d)}$, at
least outside some compact set, then the algorithm is polynomially
ergodic with rate $r/2$.
\item[(3)]
If $\exists r>0$ and $\eta\in(0,2)$ such that
$\pi(\mathbf{x}) \propto\break \| \mathbf{x} \|_2^{-(r+d)}$, at
least for large enough $\mathbf{x}$, and the proposal has tails
$q(\mathbf{x})
\propto\| \mathbf{x} \|_2^{-(d+\eta)}$,
then the algorithm is polynomially ergodic with rate $r/\eta$.
\end{enumerate}
Thus posterior distributions with exponential or\break lighter tails
lead to a geometrically ergodic Markov chain, whereas polynomially
tailed posteriors can lead to polynomially ergodic chains, and even
this is only guaranteed if the tails of the proposal are at least as
heavy as the tails of the posterior. However, by using a proposal with
tails so heavy that it has infinite variance, the polynomial
convergence rate can be made as large as is desired.

\begin{algorithm}[(BlkShpCau)]\label{algo4} Our fourth algorithm is identical to
BlkShp but samples the proposed jump
from the heavy-tailed multivariate\break Cauchy. Proposals are
generated by simulating
$\bmV\sim N(\bmzero,\hat{\bmSigma})$ and $Z \sim N(0,1)$ and setting
$\bmY^*=\bmV/Z$. No acceptance rate criteria exist for proposals with
infinite variance and so the optimal scaling parameter for this
algorithm was found (for each dataset and $\hat{\bmSigma}$)
by repeating several small runs
with different scale parameters and noting which produced the best
ACT's for each dataset.
\end{algorithm}

\begin{algorithm}[(BlkShpMul)]\label{algo5} The fifth algorithm relies on the
fact that taking
logarithms of parameters shifts mass from the tails to the center of the
distribution. It uses a random
walk on the posterior of $\tilde{\theta}:=(\log\psi_1, \log
\psi_2, \log q_{12}, \log q_{21})$. Shape matrices
$\tilde{\bmSigma}$ were estimated as for Algorithm~\ref{algo3}, but using the
logarithms of the posterior output from Algorithm~\ref{algo1}.
In the original parameter space this
algorithm is equivalent to a proposal with components
$X^*_i = X_i e^{Y^*_i}$ and so has been called
the \textit{multiplicative random walk}
(see, e.g., Dellaportas and Roberts, \citeyear{DellaportasRoberts2003}).
In the original parameter space the acceptance probability is
\[
\alpha(\mathbf{x},\mathbf{x}^*) =
\min\biggl(1, \frac{\prod_1^d{x^*_i}}{\prod_1^d{x_i}}
\frac{\pi(\mathbf{x}^*)}{\pi(\mathbf{x})} \biggr).
\]
Since the algorithm is simply an additive random walk on the log
parameter space, the usual acceptance rate optimality criteria
apply.
\end{algorithm}

A logarithmic transformation is clearly only appropriate for positive
parameters and can in fact lead to a heavy left-hand tail if a
parameter (in the original space) has too much mass close to zero.
The transformation
$\tilde{\theta}_i = \operatorname{sign}(\theta_i) \log(1+\vert
\theta_i \vert)$
circumvents both of these problems.

\subsection{Additional Strategies}
Scaling and shaping of the proposal, the choice of proposal
distribution (here Gaussian or Cauchy), and an informed choice between
RWM and\break
Metropolis-within-Gibbs updates can all lead to a more efficient algorithm.
Building on these possibilities,
we now consider two further mechanisms for improving efficiency:
adaptive MCMC, and utilizing problem-specific knowledge.

\subsubsection{Adaptive MCMC}

\mbox{}

\textit{Intuition:} Algorithm~\ref{algo3} used the output from a previous MCMC
run to estimate the shape Matrix $\bmSigma$. An overall scaling
parameter was then varied to give an acceptance rate of around
$0.3$. With adaptive MCMC a single chain is run, and this chain
gradually alters its own proposal distribution
(e.g., changing $\bmSigma$), by learning about the posterior
\textit{from its own output}. This
simple idea has a major potential pitfall, however.

If the algorithm is
started away from the main posterior mass, for example in a tail or a
minor mode, then
it initially learns about that region. It therefore alters the
proposal so that it efficiently explores this region of minor
importance. Worse, in so altering the proposal the algorithm may
become even less efficient at finding the main posterior mass,
remain in an unimportant region for longer, and become even more
influenced by that unimportant region.
Since the transition kernel is
continually changing, potentially with this positive feedback
mechanism,
it is no longer guaranteed that the
overall stationary distribution of the chain is $\pi(\cdot)$.

A simple solution is
so-called \textit{finite adaptation}\break wherein
the algorithm is only allowed to evolve for the first $n_0$
iterations, after which time the transition kernel is fixed. Such
a scheme is equivalent to running a shorter ``tuning'' chain and then
a longer subsequent chain (e.g., Algorithm~\ref{algo3}). If the tuning portion of
the chain has only explored a minor mode or a tail, this still
leads to an inefficient algorithm. We would prefer to allow the chain
to eventually correct for any errors made at early iterations and yet
still lead to the intended stationary distribution. It
seems sensible that this might be achieved provided changes to the
kernel become smaller and smaller as the algorithm proceeds and
provided the above-mentioned
positive feedback mechanism can never pervert the entire algorithm.

\textit{Theory:}
At the $n$th
iteration let $\Gamma_n$ represent the choice of transition kernel;
for the RWM it might represent the current shape matrix $\bmSigma$ and
the overall scaling $\lambda$. Denote the corresponding transition kernel
$P_{\Gamma_n}(\mathbf{x},\cdot)$.
Roberts and Rosenthal (\citeyear{RobertsRosenthal2007}) derived two conditions which
together guarantee convergence to the stationary distribution. A key
concept is that of \textit{diminishing adaptation}, wherein\break changes to
the kernel must become vanishingly small as $n \rightarrow\infty$,
\[
\sup_{\mathbf{x}}\| P_{\Gamma_{n+1}}(\mathbf{x},\cdot )-P_{\Gamma
_{n}}(\mathbf{x},\cdot) \|_1
   \cip0 \quad  \mbox{as } n\rightarrow\infty.
\]
A second \textit{containment} condition
considers the $\epsilon$-convergence time under
repeated application of a fixed kernel, $\gamma$, and starting point
$\mathbf{x}$,
\[
M_\epsilon(\mathbf{x},\gamma) := \inf_n \{n \ge1\dvtx
\| P_\gamma^n(\mathbf{x},\cdot)-\pi(\cdot) \|_1\le\epsilon
\},
\]
and requires that for all $\delta>0$ there is an $N$ such that for all $n$
\[
\mathbb{P}\bigl(M_\epsilon(\bmX_n,\Gamma_n)\le N | \bmX_0=\mathbf
{x}_0,\Gamma_0=\gamma_0\bigr) \ge1-\delta.
\]
The containment condition is difficult to check
in practice;
some criteria are provided in the work of Bai, Roberts and Rosenthal (\citeyear{BaiRobertsRosenthal2009}).

Adaptive MCMC is a highly active research area and so we
confine ourselves to an adaptive version of Algorithm~\ref{algo5}.
Roberts and Rosenthal (\citeyear{RobertsRosenthal2009a}) described an adaptive RWM algorithm for which
the proposal at the $n$th iteration is sampled from a mixture of an
adaptive $N (\bmzero,\frac{1}{d}2.38^2\hat{\bmSigma}_n )$
and a nonadaptive Gaussian
distribution; here $\hat{\bmSigma}_n$ is the variance matrix
calculated from the previous
$n-1$ iterations of the scheme. Changes to the variance matrix are
$O(1/n)$ at the $n$th iteration and so the algorithm satisfies
the diminishing adaptation condition.

Choice of the overall scaling factor
$2.38^2/d$ follows directly from the optimal
scaling limit results reviewed in Section
\ref{sect.optimal.scale}, with $J=1$ or $\kxd=\kyd$. In general,
therefore, a different scaling might be appropriate, and so
our scheme extends that of Roberts and Rosenthal (\citeyear{RobertsRosenthal2009a})
by allowing the overall scaling factor to adapt.



\begin{algorithm}[(BlkAdpMul)]\label{algo6} Our adaptive\break  \mbox{MCMC} algorithm is a block
multiplicative random walk which samples jump
proposals on the\break log-posterior from the mixture
\[
\bmY\sim\cases{\displaystyle
N (\bmzero,m_n^2\tilde{\bmSigma}_n )& w.p.
$1-\delta$,\vspace*{2pt}\cr\displaystyle
N \biggl(\bmzero,\frac{1}{d}\lambda_0^2\bmI\biggr)& w.p.
$\delta$.
}
\]
Here $\delta=0.05$, $d=4$, and $\tilde{\bmSigma}_n$ is the variance
matrix of
the logarithms
of the posterior sample to date. A few minutes were spent tuning the
block multiplicative random walk with proposal variance $\frac
{1}{4}\lambda_0^2
\bmI$ to give at least a reasonable value for $\lambda_0$
(acceptance rate $\approx0.3$), although this is not strictly necessary.
\end{algorithm}

To ensure a sensible nonsingular $\tilde{\bmSigma}_n$,
proposals from the adaptive part of the mixture were only allowed once
there had been at least 10 proposed jumps accepted.
The overall scaling factor for the adaptive part of the kernel, $m_n$, was
initialized to $m_0=2.38/d^{1/2}$ and an adaptation
quantity $\Delta=\break m_0/100$ was defined.
If iteration $i$ was from the nonadaptive part of the kernel, then
$m_{i+1}\leftarrow m_i$; otherwise:
\begin{itemize}
\item
If the proposal was rejected, then $m_{i+1}\leftarrow m_i-\Delta/i^{1/2}$.
\item
If the proposal was accepted, then $m_{i+1}\leftarrow m_i+2.3 \Delta/i^{1/2}$.
\end{itemize}
This leads to an equilibrium acceptance rate of\break $1/3.3\approx30\%$,
the target acceptance rate for the other block updating algorithms
which use Gaussian proposals (Algorithms~\ref{algo1}, \ref{algo3}, and~\ref{algo5}). Changes to
$m$ are scaled by $i^{1/2}$ since
they must be large enough to adapt to changes in
the covariance matrix yet small enough that an equilibrium value is
established relatively quickly. As with the variance matrix,
such a value would then only change
noticeably if there were consistent evidence that it should.

\subsubsection{Utilizing problem-specific knowledge}
\label{sect.prob.spec}

\mbox{}

\textit{Intuition:} Algorithms are always applied to
specific datasets with specific forms for the likelihood and
prior. Combining techniques such as
optimal scaling and shape adjustment with problem-specific knowledge
can often markedly improve efficiency. In
the case of the MMPP we define a reparameterization based on the
intuition that for an MMPP with $\psi_1 \approx\psi_2$ (as in D2)
the data
contain a great deal of information about the average intensity but relatively
little information about the difference between the intensities.

%
\begin{table*}[b]
\caption{Summary of the algorithms used in this paper}
\label{table.algorithms}
\begin{tabular}{@{}lll@{}}
\hline
\textbf{No.} & \textbf{Abbreviation} & \textbf{Description}\\
\hline
\phantom{1}1&Blk &
Block additive with tuned proposal $N(\bmzero,\lambda^2\bmI)$.\\
\phantom{1}2&MwG &
Sequential additive with tuned proposals
$N(\bmzero,\lambda_i^2)$ ($i=1,\ldots, 4$).\\
\phantom{1}3&BlkShp &
Block additive with tuned proposal $N(\bmzero,\lambda^2\hat{\bmSigma
})$.\\
\phantom{1}4&BlkShpCau &
Block additive with tuned proposal $\operatorname{Cauchy}(\bmzero
,\lambda
^2\hat{\bmSigma})$.\\
\phantom{1}5&BlkShpMul &
Block multiplicative with tuned proposal $N(\bmzero,\lambda^2\tilde
{\bmSigma})$.\\
[3pt]
\phantom{1}6&BlkAdpMul &
Block multiplicative with adaptively tuned mixture proposal.\\
\phantom{1}7&MwGRep &Sequential multiplicative/additive
Gaussian; reparameterization.\\
\phantom{1}8&MwGRepCau &Sequential multiplicative Gaussian and additive
Cauchy; reparameterization.\\
[3pt]
\phantom{1}9&IndShp& Block independence sampler with tuned proposal $t_5(0,\hat
{\bmSigma})$.\\
10&Gibbs& Hidden data Gibbs sampler of Fearnhead and Sherlock
(\citeyear{FearnheadSherlock2006}).\\
\hline
\end{tabular}
\end{table*}

\textit{Theory:}
For a two-dimensional MMPP
define an
overall transition intensity,
stationary distribution,\break  mean intensity at stationarity,
and a measure of the difference between
the two event intensities as follows:
%
\begin{eqnarray}
\label{eqn.reparam.a}
q &:=& q_{12} + q_{21}
, \quad
\bmnu:= \frac{1}{q} [q_{21},q_{12} ]
,\nonumber
\\[-8pt]
\\[-8pt]
\psibar&:=& \bmnu^t \bolds{\psi}
 \quad \mbox{and} \quad
\delta:=\frac{(\psi_2-\psi_1)}{\psibar.}\nonumber
\end{eqnarray}
Let $t_{\mathrm{obs}}$ be the total observation time and $\bmt$ the vector of
observed event times.
If the Poisson event intensities are similar, $\delta$ is small, and
Taylor expansion of the log-likelihood
in $\delta$ (see Sherlock, \citeyear{Sherlock2006}) gives
%
\begin{eqnarray}\label{eqn.firstexpansion}
&&l(\psibar, q, \delta, \nu_1)\nonumber\\
&& \quad =
n \log\psibar- \psibar t_{\mathrm{obs}}
+ 2 \delta^2 \nu_1 \nu_2 f(\psibar\bmt, q \bmt)
\\
&& \qquad {}+ \delta^3 \nu_1 \nu_2 (\nu_2 - \nu_1) g(\psibar\bmt, q \bmt)
+ O(\delta^4)\nonumber
\end{eqnarray}
for some $f(\cdot,\cdot)$ and $g(\cdot,\cdot)$. Consider a
reparameterization from
$(\psi_1,\psi_2,q_{12},q_{21})$ to $(\psibar,q,\alpha,\beta)$ with
%
\begin{equation}
\alpha:= 2 \delta(\nu_1\nu_2 )^{1/2}
 \quad \mbox{and} \quad
\beta:= \delta(\nu_2-\nu_1).
\end{equation}
Parameters $\psibar$; $q$ and $\alpha$; and $\beta$ (in this order) capture
decreasing amounts of variation in
the log-likelihood and so, conversely, it might be anticipated that
there be corresponding decreasing
amounts of
information about these parameters contained in the likelihood. Hence
very different scalings might be required for each.

\begin{algorithm}[(MwGRep)]\label{algo7} A Metropolis-within-Gibbs update
scheme was
applied to the reparameterization
$(\psibar,q,\alpha,\beta)$. A multiplicative random walk
was used for each of the first three parameters (since they are positive)
and an additive update was used for $\beta$.
Scalings for each of the four parameters were
chosen to give acceptance rates of between
$0.4$ and $0.45$.
\end{algorithm}

\begin{algorithm}[(MwGRepCau)]\label{algo8} Our final algorithm is identical to
MwGRep except that
additive updates for $\beta$ are proposed from a
\textit{Cauchy} distribution. The Cauchy scaling was optimized to give
the best ACT over the first 1000 iterations.
\end{algorithm}

\section{Results}
\label{sect.results}
The eight algorithms described in Section \ref{sect.theory.practice}
are summarized in
Table \ref{table.algorithms}. The table includes two further
\mbox{algorithms}, an independence
sampler (Algorithm 9: IndShp), and the Gibbs
sampler of Fearnhead and Sherlock (\citeyear{FearnheadSherlock2006}) (Algorithm 10: Gibbs); these
were included to benchmark the efficiency of RWM
algorithms against some sensible alternatives. The independence
sampler used a multivariate $t$ distribution with five degrees of freedom
and the same set of covariance matrices as Algorithm~\ref{algo3}.

Each RWM variation was tested against datasets D1
and D2 as described in Section \ref{sect.mmpp.data}.
For each dataset, each algorithm was started from the known ``true''
parameter values
and was run three times with three different random seeds (referred to as
Replicates 1--3).
All algorithms were run for
11,000 iterations; a burn in of 1000 iterations was
sufficient in all cases.

Priors were independent and exponential with\break means the known ``true''
parameter values.
The likelihood of an MMPP with maximum and minimum Poisson intensities
$\psi_{\max}$ and $\psi_{\min}$ and with
$n$ events observed over a time window
of length $t_{\mathrm{obs}}$ is bounded above by
$\psi_{\max}^n e^{-\psi_{\min} t_{\mathrm{obs}}}$.
In this article only MMPP parameters and their logarithms are
considered for estimation.
Since exponential priors are employed the parameters and their
logarithms therefore have finite variance,
and geometric ergodicity is guaranteed.

\begin{figure*}

\includegraphics{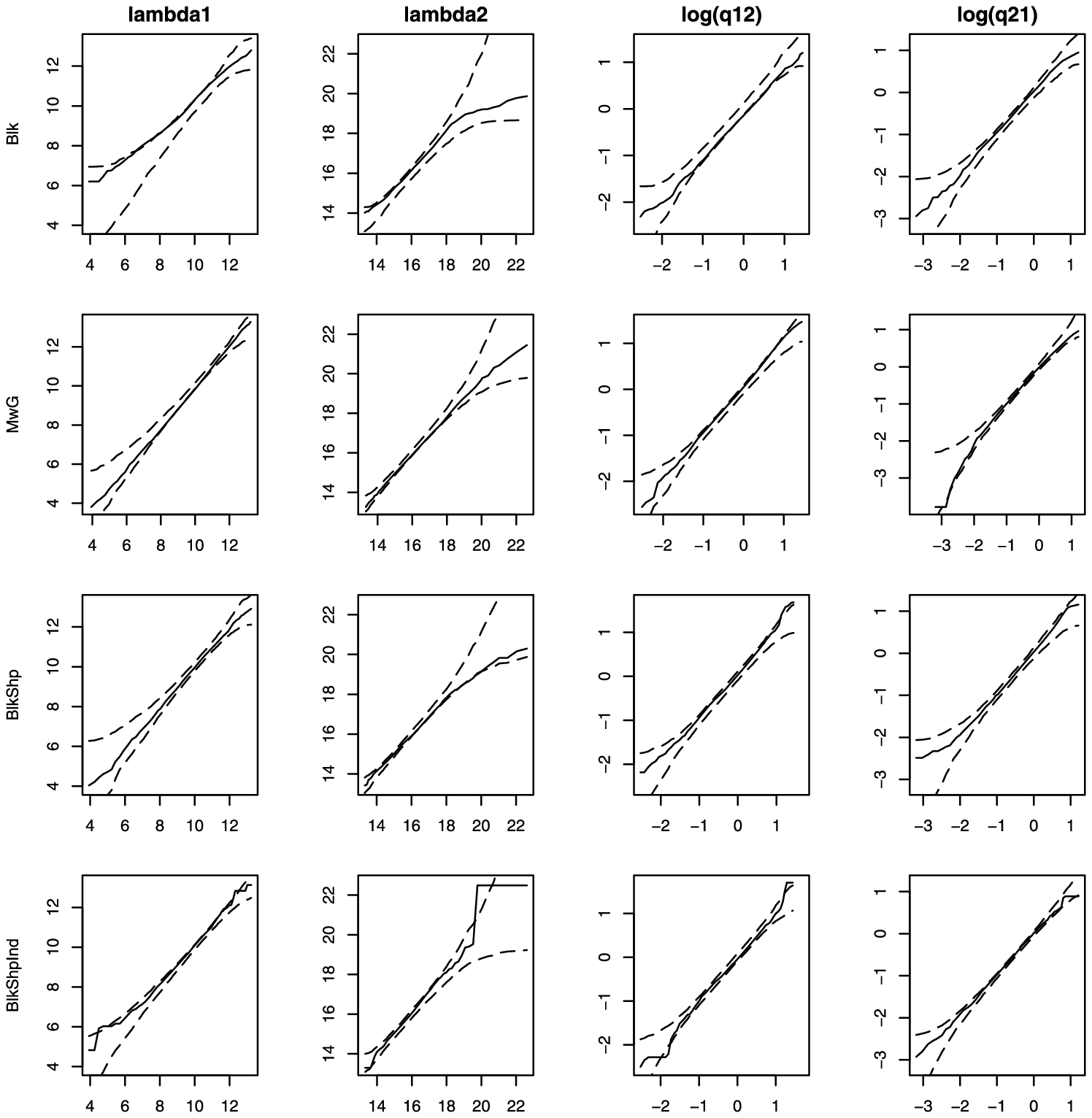}

\caption{QQ plots for algorithms Blk, MwG, BlkShp, and IndShp,
on \textup{D}2
(Replicate 1).
Dashed lines are approximate 95\% confidence
limits obtained by repeated sampling from iterations 1000 to
100,000 of a Gibbs sampler run; sample sizes were 10,000/ACT,
which is the
effective sample size of the data being compared to the Gibbs
run.}
\label{fig.qq.3.1}
\end{figure*}

The accuracy of posterior simulations is
assessed via QQ plot comparison with the output from a very long run of
a Gibbs sampler (see Section \ref{sect.accuracy}).
QQ plots for almost all replicates
were almost entirely within their 95\%
confidence bounds.
Figure \ref{fig.qq.3.1} shows such plots for
Algorithms~\ref{algo1}--\ref{algo3} and 9 (the independence sampler) on dataset D2
(Replicate 1).
In general these combinations produced the least accurate
performance, and only with the independence sampler is
there reason to doubt that the posterior sample is a reasonable
representation of the true posterior. The relatively poor performance
on D2
of Algorithms~\ref{algo1}--\ref{algo3} and especially Algorithm 9 is repeated for
the other two replicates.
The
third replicate of Algorithm~\ref{algo4} on D2 also showed an imperfect fit in the
tails.

The integrated ACT was estimated for each parameter and each replicate
using the final 10,000 iterations from that replicate.
Calculation of the likelihood is by far the most computationally
intensive operation (taking approximately $99.8\%$ of the total CPU
time) and is performed four times for each
Metropolis-within-Gibbs-iteration (once for each parameter)
and only once for each block update; a similar calculation is
performed once for each update of the Gibbs sampler. To give a
truer indication of overall efficiency the ACTs for each
Metropolis-within-Gibbs replicate have therefore been multiplied by 4.
Table \ref{table.acf.one} shows the mean adjusted ACT
for each algorithm, parameter, and dataset.
For each set of three replicates most of the ACTs lay within 20\% of
their mean, and for
the exceptions (Blk and BlkShpCau for datasets D1 and D2, and BlkShp
and BlkShpMul for dataset D2) full sets of ACTs
are given in Table \ref{table.acf.two}
in the \hyperref[sect.extra.act]{Appendix}.

In general all algorithms performed better on D1 than on D2 because,
as discussed in Section \ref{sect.mmpp.data},
dataset D1 contains more information on
the parameters than D2; it therefore has lighter tails and is more easily
explored by the chain.

\begin{table*}
\tablewidth=\textwidth
\caption{Mean estimated integrated autocorrelation time for the four
parameters over three independent replicates for datasets \textup{D}1 and \textup{D}2}\label{table.acf.one}
\begin{tabular*}{\textwidth}{@{\extracolsep{\fill}}ld{2.1}d{2.1}d{2.1}d{2.1}d{3.0}d{3.0}d{3.0}c@{}}
\hline
&   \multicolumn{4}{c}{\textbf{D1}} &  \multicolumn{4}{c@{}}{\textbf{D2}}   \\
\ccline{2-5,6-9}
\textbf{Algorithm} & \multicolumn{1}{c}{$\bolds{\psi_1}$} & \multicolumn{1}{c}{$\bolds{\psi_2}$} & \multicolumn{1}{c}{$\bolds{\log(q_{12} )}$} &
\multicolumn{1}{c}{$\bolds{\log(q_{21} )}$}
& \multicolumn{1}{c}{$\bolds{\psi_1}$} & \multicolumn{1}{c}{$\bolds{\psi_2}$} & \multicolumn{1}{c}{$\bolds{\log(q_{12} )}$} &
\multicolumn{1}{c@{}}{$\bolds{\log(q_{21} )}$}\\
\hline
Blk & 66 & 126 & 15 & 19 & 176 & 175 & 80 & 70\\
MwG\tabnoteref[*]{tt1} & 22 & 22 & 33 & 33 & 103 & 90 & 114& 99\\
BlkShp & 13 & 18 & 13 & 15 & 46 & 25 & 37 & 36\\
BlkShpCau & 19 & 32 & 25 & 24 & 63 & 50 & 56 & 38\\
BlkShpMul & 13 & 17 & 13 & 15 & 33 & 26 & 22 & 16\\
[3pt]
BlkAdpMul & 12 & 12 & 14 & 14 & 20 & 20 & 17 & 23\\
MwGRep\tabnoteref[*]{tt1} & 13 & 14 & 32 & 44 & 20 & 23 & 23 & 21\\
MwGRepCau\tabnoteref[*]{tt1}& 14 & 15 & 37 & 42 & 24 & 233 & 25 & 23 \\
[3pt]
IndShp$^+$ & 3.7 & 5.5 & 3.5 & 3.7 & & & & \\
Gibbs & 4.2 & 3.2 & 5.7 & 5.9 & 26 & 19 & 32 & 27\\
\hline
\end{tabular*}
\tabnotetext[]{tt1}{\textit{Notes}: $^*$Estimates for MwG replicates have been multiplied by 4 to
provide figures comparable with full block updates in terms of CPU
time. $^+$ACT results for the independence sampler for D2 are
irrelevant since the MCMC sample was not an accurate
representation of the~posterior.}
\end{table*}

The simple block additive algorithm using Gaussian proposals
with variance matrix proportional to the identity matrix (Blk)
performs relatively poorly on both datasets. In absolute terms there
is much less uncertainty about the transition intensities
$q_{12}$ and $q_{21}$ (both are close to
$1$) than in the Poisson intensities
$\psi_1$ ($10$) and $\psi_2$ ($17$ for D1 and $30$ for D2) since
the variance of the output from a Poisson process is proportional to
its value. The optimal single-scale parameter necessarily tunes to the
smallest variance and hence explores $q_{12}$ and $q_{21}$ much more
efficiently than $\psi_1$ and $\psi_2$.

Overall performance improves enormously
once\break block proposals are from a Gaussian with approximately the
correct shape (BlkShp).
The efficiency of the Metropolis-within-Gibbs algorithm with
additive Gaussian updates (MwG) lies somewhere between the efficiencies of
Blk and BlkShp but the improvement over Blk is larger
for dataset D1 than for dataset D2. As discussed in
Section \ref{sect.mmpp.data} the parameters in D1 are more nearly
independent than the parameters in D2. Thus for dataset D1 the
principal axes of an elliptical approximation to the posterior are
more nearly parallel to the cartesian axes. Metropolis-within-Gibbs
updates are
(by definition) parallel to each of the cartesian axes and so can make
large updates almost directly along the major axis of the ellipse for
dataset D1.

For the heavy-tailed posterior of dataset D2
we would expect block
updates resulting from a Cauchy proposal (BlkShpCau) to be more
efficient than
those from a Gaussian proposal. However, for both datasets Cauchy
proposals are slightly less efficient than Gauss\-ian proposals.
It is likely that the
heaviness of the Cauchy tails leads to more proposals with at least
one negative parameter, such proposals being automatically
rejected. Moreover, $\hat{\bmSigma}$ represents the main posterior mass,
yet some large Cauchy jump proposals from this mass will be in
the posterior tail. It may be that $\hat{\bmSigma}$ does not
accurately represent the shape of the posterior tails.

Multiplicative updates (BlkShpMul) make little difference for D1, but
for the relatively heavy-tailed D2 there is a definite
improvement over BlkShp. The
adaptive multiplicative algorithm (BlkAdpMul) is slightly more
efficient still, since the estimated variance
matrix and the overall scaling are refined throughout the run.

As was noted earlier in this section, due to our choice of exponential
priors the quantities estimated in
this article have exponential or lighter posterior tails and so
all the nonadaptive algorithms in this article are geometrically
ergodic. The theory in Section
\ref{sect.tail.explore} suggests ways to improve tail exploration for
polynomially ergodic algorithms and so, strictly speaking, need not
apply here. However, the exponential decay only becomes
dominant some distance from the posterior mass, especially for dataset
D2. Polynomially
increasing terms in the likelihood ensure that initial decay is
slower than exponential, and that the multiplicative random walk is
therefore more efficient than the additive random walk.

\begin{figure}

\includegraphics{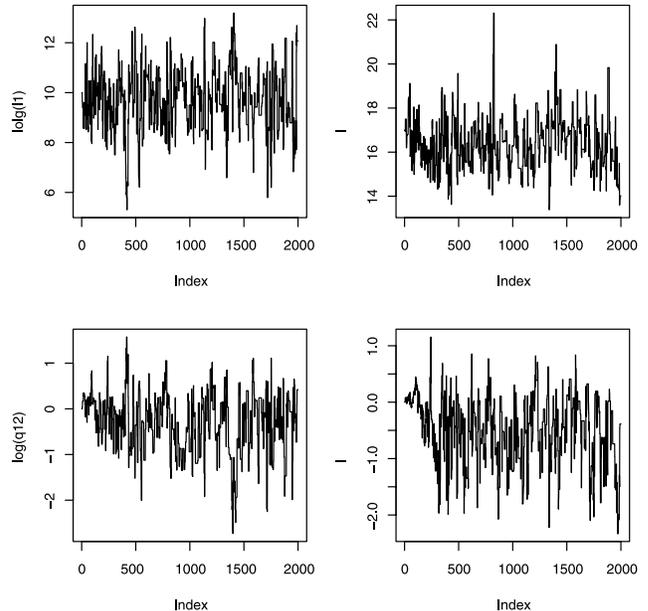}

\caption{Traceplots for the first 2000 iterations
of BlkAdpMul on dataset \textup{D}2 (Replicate 1).}
\label{fig.adp.trace}
\end{figure}

The adaptive overall scaling $m$ showed variability of $O(0.1)$ over
the first 1000 iterations after which time it quickly settled down to
$1.2$ for all three replicates on D1 and to $1.1$ for all three
replicates on
D2. Both of these values are very close to the scaling of $1.19$ that
would be used for a four-dimensional update in the scheme of
Roberts and Rosenthal (\citeyear{RobertsRosenthal2009a}). The algorithm similarly learned very
quickly about the
variance matrix $\bmSigma$, with individual terms settling down after
less than 2000 iterations, and with exploration close to optimal
after less than 500 iterations. This can be seen clearly in
Figure \ref{fig.adp.trace}
which shows traceplots for the first 2000 iterations of
the first replicate of BlkAdpMul on D2.

The adaptive algorithm uses its own history to learn
about $d(d+1)/2$ covariance terms and a best overall scaling.
One would therefore expect that the larger the number of
parameters, $d$, the more iterations are required for the
scheme to learn about all of the adaptive terms and hence
reach a close to optimal efficiency. To test this a dataset (D3) was
simulated from a three-dimensional MMPP with $\bolds\psi
=[10,17,30]^t$ and
$q_{12}=q_{13}=q_{21}=q_{23}=q_{31}=q_{32}=0.5$. The following adaptive
algorithm was then run three times, each for 20,000 iterations.

\renewcommand{\thealgorithm}{6b}
\begin{algorithm}[{[BlkAdpMul(b)]}]\label{algo6b} This adaptive algorithm is
identical to BlkAdpMul (with $d=9$) except that no adaptive proposals
were used until at least 100 nonadaptive proposals had been
accepted, and that if an adaptive proposal was
accepted then the overall scaling was updated with
$m\leftarrow m+3 \Delta/i^{1/2}$ so that the equilibrium
acceptance rate was approximately $0.25$.
\end{algorithm}

\begin{figure}

\includegraphics{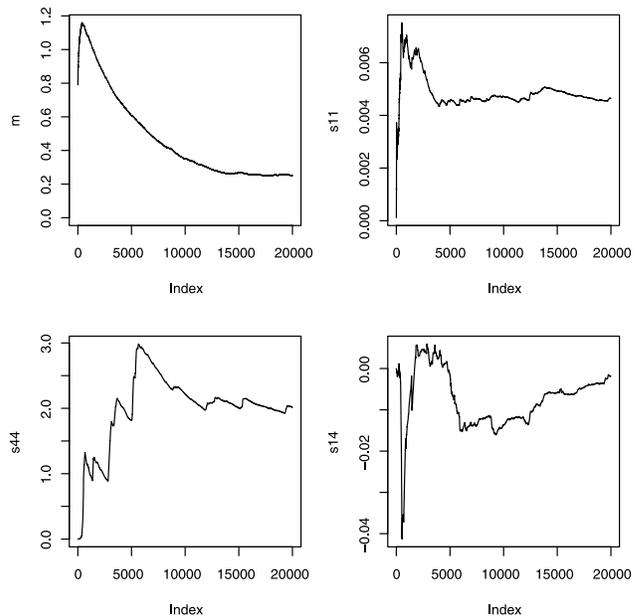}

\caption{Plots of the adaptive scaling parameter $m$ and three
estimated covariance parameters $\operatorname{Var}[\psi_1]$,
$\operatorname{Var}[q_{12}]$,
and $\operatorname{Cov}[\psi_1,q_{12}]$ for BlkAdpMul(b)
on dataset D3 (Replicate 1).}
\label{fig.adp.3D.adp.parms}
\end{figure}

Figure \ref{fig.adp.3D.adp.parms}
shows the evolution of four of the 46 adaptive
parameters (Replicate 1). All parameters seem
close to their optimal values after 10,000 iterations, although
covariance parameters appear to be still slowly evolving even after
20,000 iterations. In contrast, traceplots of parameters (not shown)
reveal that the speed of exploration of the posterior is close to its final
optimum after only 1500 iterations.
This behavior was repeated across the other two replicates, indicating
that, as with the two-dimensional adaptive and nonadaptive runs,\break
even a very rough approximation to the variance matrix improves
efficiency considerably. Over the full 20,000 iterations, all three replicates
showed a definite multimodality with $\lambda_2$
often close to either $\lambda_1$ or $\lambda_3$,
indicating that the data might reasonably be explained by a
two-dimensional MMPP. In all three replicates the optimal
scaling settled between $0.25$ and $0.3$,
noticeably lower than the Roberts and Rosenthal (\citeyear{RobertsRosenthal2009a})
value of $2.38/\sqrt{9}$. With
reference to Section \ref{sect.optimal.scale}
this is almost certainly due to the
roughness inherent in a multimodal posterior.

The reparameterization of Section \ref{sect.prob.spec}
was designed for datasets
similar to D2, and on this dataset
the resulting Metropolis-within-Gibbs algorithm\break (MwGRep) is
at least as efficient as the adaptive multiplicative random walk. On
dataset D1, however, exploration of $q_{12}$ and $q_{21}$ is arguably
less efficient than for the Metropolis-within-Gibbs algorithm with the
original parameter set. The lack of improvement when using a Cauchy
proposal for $\beta$ (MwGRepCau) suggests that this inefficiency is
not due to
poor exploration of the potentially heavy-tailed $\beta$.
Further investigation in the $(\psibar,q,\alpha,\beta)$
parameter space showed that for dataset D1
only $q$ was explored efficiently; the
posteriors of $\psibar$ and $\beta$ were strongly positively
correlated ($\rho\approx0.8$), and both $\psibar$ and $\beta$ were
strongly negatively
correlated with $\alpha$ ($\rho\approx-0.65$). Posterior
correlations were small $\vert\rho \vert<0.3$ for all parameters with
dataset D2 and for all correlations involving $q$ for dataset D$1$.

The optimal scaling for
the one-dimensional additive Cauchy proposal in MwGRepCau was
approximately two thirds of the optimal scaling for the
one-dimensional additive Gaussian proposal in MwGRep. In four
dimensions the ratio was approximately one half. These ratios allow
the Cauchy proposals to produce
similar numbers of small to medium sized jumps to the Gaussian proposals.

The independence sampler is arguably the most efficient of all of the
algorithms considered for D1. However,
as discussed earlier in this section, there are doubts about the
accuracy of its exploration of D2.
Mengersen and Tweedie (\citeyear{MengersenTweedie1996}) showed that an independence sampler is uniformly
ergodic if and only if the ratio of the proposal density to the target
density is bounded below, and that one minus this ratio gives the
geometric rate of
convergence. To ensure the lower bound it
is advisable to propose from a relatively heavy-tailed distribution,
such as the $t_5$ used here. The problem in this instance arises
because dataset D2 could, just possibly, have been generated by
a single Poisson process with intensity $\psi\approx
(\psi_1+\psi_2)/2$. The resulting minor mode (or, more precisely, ridge)
is some distance from the
center of the distribution, resulting in a low ratio of proposal and
target densities.

The Gibbs sampler of Fearnhead and Sherlock (\citeyear{FearnheadSherlock2006}) is accurate,
with its efficiency directly related to the amount of
information about the hidden Markov chain that is available from the data
(Sherlock, \citeyear{Sherlock2006}). Thus for D1 the Gibbs sampler is more efficient
than the best RWM algorithms, but this is not the case for D2.

\section{Discussion}
\label{sect.discussion}
We have described the theory and intuition behind a number of
techniques for improving the efficiency of random walk Metropolis
algorithms and tested these on two data sets generated from Markov
modulated Poisson processes (MMPPs). Tests on these datasets also
showed a sensibly implemented RWM to be at least as good as some of
the other available MCMC algorithms.
Some RWM implementations were uniformly successful at
improving efficiency, while for others success depended on the shape
and/or tails of the posterior. All of the underlying concepts
discussed here are quite general and easily applied to statistical
models other than the MMPP.

Simple acceptance rate
tuning to obtain the optimal overall variance term
for a symmetric Gaussian proposal can increase efficiency by many
orders of
magnitude. However, with our datasets, even after such tuning, the
RWM algorithm was very inefficient. The effectiveness of the sampling
increased enormously once the shape of the posterior was taken into
account by proposing from a Gaussian with variance proportional to an
estimate of the posterior variance. For Algorithms~\ref{algo3},  \ref{algo4}, and~\ref{algo5}
the posterior variance was estimated through a short ``training
run''---the first 1000 iterations after burn in of Algorithm~\ref{algo1}.

As expected, use of the ``multiplicative random walk'' (Algorithm~\ref{algo5}),
a random walk on the
posterior of the logarithm of the parameters, improved efficiency most
noticeably on the posterior with the heavier tails. However, contrary
to expectation, even on the
heavier tailed posterior an additive
Cauchy proposal (Algorithm~\ref{algo4}) was, if anything, less efficient than a
Gaussian. Tuning of Cauchy proposals was also more time-consuming
since simple acceptance rate criteria could not be used.

Algorithm~\ref{algo6} combined the successful strategies of optimal scaling,
shape tuning, and transforming the data, to create a
multiplicative random walk which learned the most efficient shape and scale
parameters from its own history as it
progressed. This
adaptive scheme was easy to implement and was arguably
the most efficient RWM for each of the datasets. A slight
variant of this algorithm was used to explore the posterior of a
three-dimensional MMPP, and showed that in higher dimensions
such algorithms take longer to discover close to optimal
values for the adaptive parameters. These runs also confirmed the
finding for the two-dimensional MMPP that RWM efficiency improves
enormously with knowledge of the posterior
variance, even if this knowledge is only approximate.
For a multimodal posterior such as that found for
the three-dimensional MMPP it might be argued that a different variance matrix
should be used for each mode. Such ``regionally adaptive'' algorithms
present additional problems, such as the definition of the different
regions, and are discussed further by Roberts and Rosenthal (\citeyear{RobertsRosenthal2009a}).

\renewcommand{\thetable}{\arabic{table}}
%
\begin{table*}
\tabcolsep=0pt
\tablewidth=345pt
\caption{Estimated ACT for the four
parameters, on three independent replicates for Blk and BlkShpCau on
dataset \textup{D}1 and Blk, BlkShp, BlkShpCau, and BlkShpMul on dataset \textup{D}2}
\label{table.acf.two}
\begin{tabular*}{345pt}{@{\extracolsep{\fill}}la{3.7}a{3.7}a{2.6}a{2.6}@{}}
\hline
\textbf{Algorithm} & \multicolumn{1}{c}{$\bolds{\psi_1}$} & \multicolumn{1}{c}{$\bolds{\psi_2}$}
& \multicolumn{1}{c}{$\bolds{\log(q_{12} )}$} &
\multicolumn{1}{c@{}}{$\bolds{\log(q_{21} )}$}\\
\hline
Blk (D1) & 59,64,75 & 120,155,104 & 12,15,17 & 19,21,17\\
BlkShpCau (D1) & 28,16,12 &36,29,31 & 20,20,35 & 26,23,24\\
Blk (D2) & 121,259,146 & 107,262,157 & 41,139,61 & 51,110,48\\
BlkShp (D2) & 54,51,34 & 23,24,29 & 40,45,27 & 50,35,23\\
BlkShpCau (D2) & 46,51,92 & 46,57,48 & 31,42,94 & 39,41,34\\
BlkShpMul (D2) & 53,24,23 & 22,33,25 & 20,23,24 & 17,18,13\\
\hline
\end{tabular*}\vspace*{2pt}
\end{table*}

Metropolis-within-Gibbs updates performed better when the
parameters were close to orthogonal, at which point the algorithms were
almost as efficient as an equivalent block updating algorithm with
tuned shape
matrix. The best Metropolis-within-Gibbs
scheme for dataset D2 arose
from a new reparameterization devised specifically for the\break
two-dimensional MMPP with parameter orthogonality in mind. On D2 this performed
nearly as well as the best scheme, the adaptive multiplicative random
walk.

The adaptive schemes discussed here
provide a significant step toward a goal of
completely automated algorithms. However,
as already discussed, for $d$ model-parameters, a posterior variance
matrix has
$O(d^2)$ components. Hence the length of any ``training run'' or of the
adaptive ``learning period'' increases
quickly with dimension. For high dimension it is therefore especially
important to utilize to the full any problem-specific knowledge that
is available so as to provide as efficient a starting algorithm as possible.

\begin{appendix}

\section*{Appendix: Runs with Highly Variable ACTs}
\label{sect.extra.act}
Three replicates were performed for each dataset and algorithm, and
ACTs are
summarized by their mean in Table \ref{table.acf.one}.
However,
for certain
combinations of the
algorithms and datasets the
ACTs varied considerably; full sets of ACTs for these replicates are given
in Table~\ref{table.acf.two}.
\end{appendix}

\end{document}